\documentclass{blois}

\usepackage{amsmath}
\usepackage{amssymb}

\providecommand{\f}[2]{\frac{{#1}}{{#2}}}

\newcommand{\ee}[1]{\begin{equation}#1\end{equation}}

\def\be{\begin{equation}}
\def\ee{\end{equation}}
\def\bea{\begin{eqnarray}}
\def\eea{\end{eqnarray}}

\newcommand{\Photo}{\includegraphics[height=35mm]{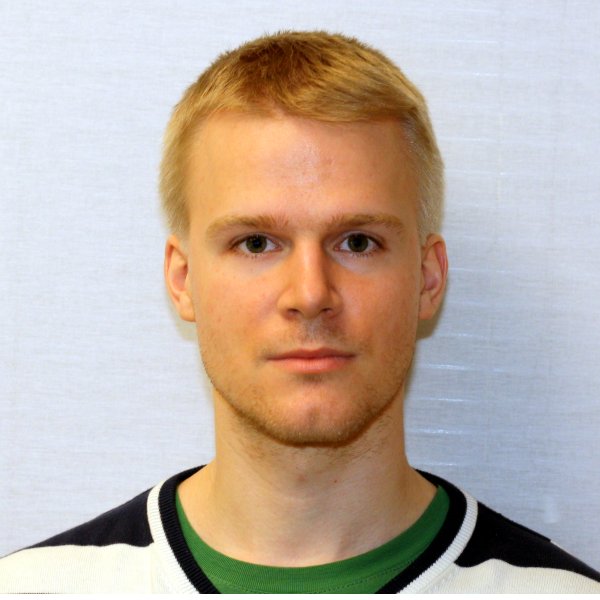}
}

\begin{document}
\vspace*{4cm}
\title{SPACETIME CURVATURE AND THE HIGGS STABILITY DURING AND AFTER INFLATION: GRAVITY TO THE RESCUE?}

\author{ T. MARKKANEN }

\address{Theoretical Physics, Blackett Laboratory, Imperial College, SW7 2AZ London, United Kingdom}

\maketitle\abstracts{We investigate the stability of the electroweak vacuum during and after inflation by taking into account the effects of classical gravity in the quantum dynamics. In particular we show that the possible instability may be avoided without any beyond the Standard Model physics. Talk presented at the \textit{27th Rencontres de Blois on Particle Physics and Cosmology}.}

\section{Introduction}
The current values for the Standard Model couplings imply a striking feature for the potential of the Higgs: at large scales there seems to exist a second minimum, with large negative energy-density. This is problematic for early universe physics, as during inflation strong background curvature may cause the Higgs to fluctuate into this second minimum, in clear contradiction with the current observation of the electroweak vacuum. Currently the most accurate calculation \cite{Buttazzo:2013uya} for the potential of the Standard Model (SM) Higgs tells us that the instability scale $\Lambda_I$ where the potential turns over to negative values lies between $10^{10}$ -- $10^{12}$GeV\,. Furthermore, from the primordial tensor bound \cite{Ade:2015xua} we know that the scale of inflation can be as high as $H\sim10^{14}$GeV. As a leading approximation, the probability density of vacuum decay during inflation scales as 
\begin{equation}
P\sim \exp\left\{-8\pi^2V_{\rm max}/(3H^4)\right \}\, ,\label{eq:p}
\end{equation}
with respect to the maximum of the potential $V_{\rm max}$ with roughly $V_{\rm max}^{1/4}\sim\Lambda_I$. Hence, a large inflationary scale, $H\gg\Lambda_I$ appears to imply probable vacuum decay during inflation \cite{Fairbairn:2014zia}. But importantly, this conclusion is based on a potential calculated on a flat background, and its validity must be carefully assessed for cases with strong background curvature, such as large scale inflation. Spacetime curvature can be incorporated into the calculation of the potential by using the standard approach of quantum field theory on a curved background and for a large $H$ it turns out to provide a stabilizing mechanism, but also can enhance the instability even further \cite{Herranen:2014cua}. The two new ingredients introduced by a curved background are the generation of the so-called non-minimal coupling $\xi$ between the SM Higgs and the scalar curvature $R$, and curvature induced renormalization group flow.

\section{Effective potential in curved space}\label{subsec:prod}
If the SM behaves as a subdominant spectator on an inflating background, we can conveniently introduce the curvature corrections by invoking the resummed Heat Kernel method \cite{Jack:1985mw}. This gives the ultraviolet (UV) portion of the modes as required by stochastic quantization \cite{Starobinsky:1994bd}, which is the most convenient way of deriving (\ref{eq:p}). In de Sitter space for the relevant degrees of freedom to 1-loop order in the 't Hooft-Landau gauge the quantum corrected, or effective, potential reads
\begin{align}
V_{\rm eff}& = -\f{1}{2}m^2\phi^2 + \f{1}{2}\xi R\phi^2 + \f{1}{4}\lambda\phi^4
+ \sum\limits_{i=1}^9 \f{n_i}{64\pi^2}M_i^4(\phi)\left[\log\f{\big|M_i^2(\phi)\big|}{\mu^2} - c_i \right]\,;\label{potential}\\M^2_i(\phi)&=\kappa_i\phi^2-\kappa' +\theta_i R\, ,\label{eq:effm}
\end{align}
where the $n_i$ count the degrees of freedom and the $M_i(\phi)$ are the effective masses. The field $\phi$ is the scalar degree of freedom of the Higgs doublet that develops an expectation value at low energies resulting in the known masses for the particles.
\begin{table}
\caption{\label{tab:contributions}The effective potential (\ref{potential}) with $W^{\pm}$, $Z^0$, top quark t, Higgs $\phi$ and the Goldstone bosons $\chi_{1,2,3}$.}
\vspace{2mm}
\begin{center}
 \begin{tabular}{|c||cccccc|}
 \hline
   $\Phi$ & $~~i$ & $~~n_i$  &$~~\kappa_i$ & $\kappa'_i$          & $~~\theta_i$   & $\quad c_i~~$ \\[1mm]\hline
   $~$ & $~~1$  & $~~2$       & $~~ g^2/4$        & $0$        & $~~{1}/{12}$     & $\quad{3}/{2}~~$ \\[1mm]
   $~W^\pm$ & $~~2$  & $~~6$       & $~~ g^2/4$        & $0$        & $~~{1}/{12}$      & $\quad{5}/{6}~~$ \\[1mm]
   $~$ & $~~3$  & $-2$      & $~~g^2/4$        & $0$        & $-{1}/{6}$      & $\quad{3}/{2}~~$ \\[1mm]\hline
   $~$ & $~~4$  & $~~1$       & $~~(g^2+g'^2)/4$ & $0$        & $~~{1}/{12}$     & $\quad{3}/{2}~~$ \\[1mm]
   $Z^0$ & $~~5$  & $~~3$       & $~~(g^2+g'^2)/4$ & $0$        & $~~{1}/{12}$      & $\quad{5}/{6}~~$ \\[1mm]
   $~$ & $~~6$  & $-1$      & $~~(g^2+g'^2)/4$ & $0$        & $-{1}/{6}$     & $\quad{3}/{2}~~$ \\[1mm]\hline
   t & $~~7$  & $-12$     & $~~ y_{\rm t}^2/2$      & $0$        & $~~{1}/{12}$     & $\quad{3}/{2}~~$ \\[1mm]\hline
   $\phi$ & $~~8$  & $~~1$       & $~~3\lambda$           & $~m^2$      & $~~\xi-{1}/{6}$  & $\quad{3}/{2}~~$
\\[1mm]\hline
   $\chi_i$ & $~~9$  & $~~3$       & $~~\lambda$            & $~m^2$      & $~~\xi-{1}/{6}$   & $\quad{3}/{2}~~$
\\[.5mm]\hline
  \end{tabular}
  \end{center}
  \end{table}
The potential in  (\ref{potential}) is not yet applicable for scales relevant for vacuum instability $\phi \geq\Lambda_I$ and the reason for this lies in the renormalization scale $\mu$: the parameters are fixed to experiments at much smaller scales than $\Lambda_I$ causing the logarithms to become large. This can be cured by using the well-known technique of renormalization group (RG) improvement, which leads to a potential with coupling constants running with the energy scale. At its core, RG improvement is a statement of $\mu$ independence, so the improved result must satisfy ${d}V_{\rm eff}/{d\mu}=0$. Unfortunately, in a perturbative result the truncated higher order terms always contain a residual $\mu$ dependence. Hence, one should choose $\mu$ in such a way that the expansion is optimized \cite{Ford:1992mv}, essentially by keeping the logarithms small. A good and frequently used choice in flat space for $\phi\gg m$ is $\mu=\phi$, which gives as the leading approximation $V_{\rm eff}\approx(\lambda(\phi)/4)\phi^4$. For a strongly curved background the same prescription forces us to make a different choice as now all effective masses contain curvature contributions. 
A good choice in curved space is then
\begin{equation}
\mu^2=\phi^2+R\,.\label{eq:mu}
\end{equation} From the above one can clearly see an important difference to the flat space result: \textit{Curvature induces running of the parameters.} Another important difference is the generation of the non-minimal term $(1/2)\xi R \phi^2$. From the SM $\beta_\xi$-function we can determine that $\xi=0$ is not an fixed point of the RG flow \cite{Herranen:2014cua}, which means that a non-zero $\xi$ will always be generated by a change in energy, such as the slow change in $H$ during inflation. It then follows that: \textit{the SM contains a non-zero $\xi$ parameter}. These two effects are  clearly visible in the leading contribution at a large scale ($\phi\gg m$) i.e. running couplings in the tree-level potential
\begin{equation}
V_{\rm eff}=\f{\xi(\mu)}{2}R\phi^2+\f{\lambda(\mu)}{4}\phi^4\, ,\label{eq:lo}
\end{equation}
where $\mu$ is chosen as in (\ref{eq:mu}). The main effects of curvature can be understood from expression (\ref{eq:lo}) and trivially for small $R$ it approaches the flat space form.
\section{Stability during inflation}
\begin{figure}
\begin{center}
\includegraphics[width=0.55\textwidth]{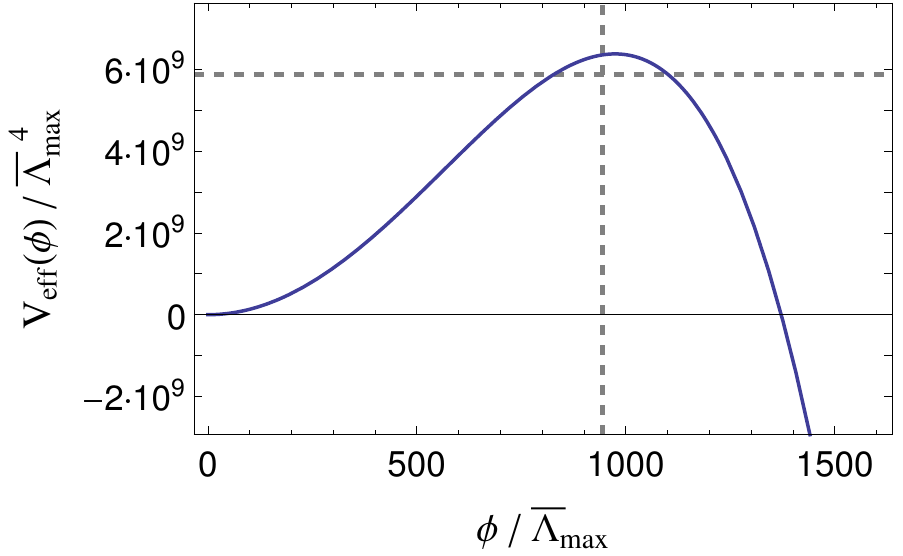}
\end{center}
\caption{\label{f}The RG improved 1-loop effective potential, with $H=10^{10}$GeV and $\xi=0.1$ at the electroweak scale.}
\end{figure}
We can now study the stability of the effective potential by using (\ref{eq:lo}) as a guide. An important observation comes by realizing that the four-point coupling is negative from the very onset of inflation if the scale of inflation $H=(R/12)^{1/2}$ exceeds the instability scale $\Lambda_I$, because of curvature induced running. This means that the only hope of having a positive potential is a large positive non-minimal term $(1/2)\xi R \phi^2$. If $\xi$ is renormalized to zero at the electroweak scale it can be shown to run to a negative value at the inflationary scale \cite{Herranen:2014cua} and hence for this choice the potential is negative and monotonically decreasing, and probably unstable. However, already when choosing $\xi$ to be $\sim 0.1$ at electroweak scales the non-minimal term runs to give a large positive contribution. As the couplings run  weakly at large scales, as a first approximation we can solve the maximum of the potential for a negative $\lambda$ but a positive $\xi$ by using (\ref{eq:lo}):
\begin{equation}
\Lambda^2_{\rm max}\approx-\frac{\xi(\mu) }{\lambda(\mu)}R\,,\qquad V_{\rm max}\approx-\frac{\left[\xi(\mu)R\right]^2}{4\lambda(\mu)}\,;\qquad \mu^2\approx R\, ,\label{eq:anapp}
\end{equation}
The above tells us that even for modest values of $\xi$ we have $V_{\rm max}\geq H^4$ and  (\ref{eq:p}) shows that the vacuum decay probability is significantly diminished. This is also visible in figure \ref{f}, where we plot the full 1-loop RG improved potential in curved space. The dashed lines correspond to the approximations in (\ref{eq:anapp}) and the $x$-axis is normalized with respect to the field value for the maximum in flat space $\overline{\Lambda}_{\rm max}$. We have chosen $H\sim10^{10}$GeV, since in the 1-loop approximation $\lambda$ becomes negative much earlier than in a the state-of-the-art derivation \cite{Buttazzo:2013uya}, roughly $\Lambda_I\sim10^8$GeV and hence for a 1-loop result this choice  corresponds to $H\gg\Lambda_I$. The result clearly shows that the peak of the potential now occurs at a scale that is $\sim 10^{3}$-times larger than $\overline{\Lambda}_{\rm max}$ and importantly that the maximum of the potential is also increased by a similar amount, approximately $V^{1/4}_{\rm max}\sim 2H$. Formula (\ref{eq:p}) then gives the transition probability $P<e^{-400}$, which indicates that $\xi \gtrsim0.1$ at the electroweak scale is enough to stabilize the potential. Note that all values for $\xi$ larger than some threshold will give a stabilizing result and in fact quickly after $\xi=1/6$ the Higgs starts behaving as a non-fluctuating massive field. The current experimental bounds for $\xi$ are very weak \cite{Atkins:2012yn} and hence the SM can be stable during inflation.
\section{Stability after inflation}
Even if there is no instability during inflation, the dynamics of the subsequent reheating phase can also trigger the instability  \cite{Herranen:2015ima}. Reheating is a complicated process and often involves non-perturbative resonance phenomena that are characterised by explosive particle production that result from the oscillations of the inflaton \cite{Kofman:1997yn}. As the SM Higgs will always be coupled to the inflaton due to the generation of $\xi$, it will feel the dynamics of reheating. The Einstein equation gives for an inflaton $\Phi$ with the potential $U(\Phi)$  in Planck units, $R=4 U(\Phi)-\dot{\Phi}^2$,
which shows that if $\Phi$ oscillates, so does $R$ and the Higgs generically has an oscillating mass $\xi R$. Importantly, the oscillations of $R$ periodically become negative, which gives rise to \textit{tachyonic resonance} that can quickly generate a large fluctuation \cite{Bassett:1997az} and possibly vacuum decay \cite{Herranen:2015ima}. In particular, reheating limits \textit{large} values of $\xi$ as the amplitude of the oscillations increases with $\xi$, which in turn makes the effect itself stronger. For $\xi\lesssim\mathcal{O}(1)$ the potential instability can be avoided \cite{Herranen:2015ima}. 
\section{Conclusions}
To conclude, gravity cannot be neglected in the quantum dynamics of the early universe vacuum instability for a large scale of inflation. The two main effects are curvature induced running of the constants and the generation of the non-minimal coupling for the Higgs. The non-minimal parameter $\xi$ can provide a stabilizing mechanism during inflation, where requiring stability results in a lower bound for $\xi$. However, in the reheating phase large values of $\xi$ result in the generation of a large fluctuation via tachyonic resonance, which can result in vacuum decay. Combining the inflationary and reheating stability limits gives at the electroweak scale \cite{Herranen:2014cua} \cite{Herranen:2015ima}
\begin{equation}
0.1\lesssim\xi\lesssim\mathcal{O}(1)
\end{equation}
\section*{Acknowledgments}
This talk is based on the research articles \cite{Herranen:2014cua} \cite{Herranen:2015ima} written in collaboration with M. Herranen, S. Nurmi and A. Rajantie. TM wishes to thank the organizers of the 27th Rencontres the Blois for a vibrant conference. This work was supported by the Osk. Huttunen Foundation.


\section*{References}

\begin{thebibliography}{99}
\bibitem{Herranen:2014cua}
  M.~Herranen, T.~Markkanen, S.~Nurmi and A.~Rajantie,
  Phys.\ Rev.\ Lett.\  {\bf 113} (2014) 21,  211102
  [arXiv:1407.3141 [hep-ph]].
  
\bibitem{Herranen:2015ima}
  M.~Herranen, T.~Markkanen, S.~Nurmi and A.~Rajantie,
  arXiv:1506.04065 [hep-ph].


\bibitem{Buttazzo:2013uya}
  D.~Buttazzo, G.~Degrassi, P.~P.~Giardino, G.~F.~Giudice, F.~Sala, A.~Salvio and A.~Strumia,
  JHEP {\bf 1312} (2013) 089
  [arXiv:1307.3536 [hep-ph]].
\bibitem{Ade:2015xua}
  P.~A.~R.~Ade {\it et al.} [Planck Collaboration],
  arXiv:1502.01589 [astro-ph.CO].

\bibitem{Fairbairn:2014zia}
  M.~Fairbairn and R.~Hogan,
  Phys.\ Rev.\ Lett.\  {\bf 112} (2014) 201801
  [arXiv:1403.6786 [hep-ph]].
  
\bibitem{Jack:1985mw}
  I.~Jack and L.~Parker,
  Phys.\ Rev.\ D {\bf 31} (1985) 2439.
  
  

\bibitem{Starobinsky:1994bd}
  A.~A.~Starobinsky and J.~Yokoyama,
  Phys.\ Rev.\ D {\bf 50} (1994) 6357
  [astro-ph/9407016].


  
\bibitem{Ford:1992mv}
  C.~Ford, D.~R.~T.~Jones, P.~W.~Stephenson and M.~B.~Einhorn,
  Nucl.\ Phys.\ B {\bf 395} (1993) 17
  [hep-lat/9210033].
  
\bibitem{Atkins:2012yn}
  M.~Atkins and X.~Calmet,
  Phys.\ Rev.\ Lett.\  {\bf 110} (2013) 5,  051301
  [arXiv:1211.0281 [hep-ph]].
  
  
\bibitem{Kofman:1997yn}
  L.~Kofman, A.~D.~Linde and A.~A.~Starobinsky,
  Phys.\ Rev.\ D {\bf 56} (1997) 3258
  [hep-ph/9704452].
  
  

\bibitem{Bassett:1997az}
  B.~A.~Bassett and S.~Liberati,
  Phys.\ Rev.\ D {\bf 58} (1998) 021302
   [Phys.\ Rev.\ D {\bf 60} (1999) 049902]
  [hep-ph/9709417].


  

\end{thebibliography}
\end{document}